\documentclass[aps,prb,reprint,nopacs,superscriptaddress,longbibliography]{revtex4-2}

\usepackage{graphicx}
\usepackage{dcolumn}
\usepackage{bm}
\usepackage{xcolor}
\usepackage{amsmath}
\usepackage{hyperref}
\usepackage{cleveref}
\usepackage{placeins}
\usepackage{float}

\newif\ifcleanversion
\cleanversiontrue
\ifcleanversion
  \colorlet{red}{black}\colorlet{cyan}{black}
\fi

\usepackage{graphicx} 
\newcommand{\br}{\mathbf{r}}
\newcommand{\brP}{\mathbf{r'}}

\newcommand{\x}{_\mathrm{x}}
\newcommand{\exx}{^\mathrm{exact}}
\newcommand{\sll}{^\mathrm{sl}}

\newcommand{\phiiP}{\phi_i(\brP)}

\newcommand{\phiiS}{\phi_i^*(\br)}

\newcommand{\uni}{^\mathrm{uni}}
\newcommand{\vW}{^\mathrm{vW}}
\newcommand{\Hyd}{^\mathrm{Hyd}}
\newcommand{\nog}{_\mathrm{0}}

\begin{document}

\preprint{new_mGGA}

\title{Meta-generalized gradient
approximation made in the Hartree gauge}%

\author{Yan Oueis}
\thanks{These authors contributed equally to this work}%
\author{Akilan Ramasamy}%
\thanks{These authors contributed equally to this work}%
\author{James W. Furness}%
\altaffiliation[Current address: ]{Schr\"odinger, Inc., New York, New York 10036, USA}%
\author{Jamin Kidd}%
\author{Timo Lebeda}%
\author{Jianwei Sun}%
 \email[Corresponding Author:]{jsun@tulane.edu}
\affiliation{%
Department of Physics and Engineering Physics, Tulane University, New Orleans, LA 70118, USA
}%

\date{\today}

\begin{abstract}


In density functional theory (DFT), exact constraints, fundamental mathematical properties of the exchange-correlation (XC) energy and its underlying XC hole, along with paradigm systems such as the uniform electron gas and the hydrogen atom have been instrumental in developing exchange-correlation (XC) density functional approximations (DFAs). However, since the spatial XC energy density is not uniquely defined, its exact constraints can only be formulated within a chosen gauge and are therefore seldom utilized in DFA construction.
Here, we propose a meta-generalized gradient approximation
for the exchange energy, explicitly constructed within the Hartree gauge, using the hydrogen atom's exchange energy density for gauge alignment in core and asymptotic regions.
{\color{red}
Constructing semilocal DFAs consistently within a single, well-defined gauge provides a foundation for nonlocal density functionals that require gauge alignment, and makes exchange energy densities from accurate reference calculations directly usable as targets for data-driven DFA development.
}

\end{abstract}


\maketitle

\section{Introduction}
\label{sec:intro}


Due to its balance of accuracy and computational efficiency, density functional theory (DFT)~\cite{hohenberg1964inhomogeneous, kohn1965self} has been widely used as a major electronic structure method across chemistry, materials science, and condensed matter physics~\cite{burke2012perspective, becke2014perspective, yu2016perspective,mardirossian2017thirty, zhang2025advances}. 
While DFT is exact in principle for ground-state properties, practical use in the Kohn-Sham (KS) scheme requires approximating the exchange-correlation (XC) energy as a functional of the electron density. Significant progress has been made in developing density functional approximations (DFAs), from local spin density approximation (LSDA)~\cite{kohn1965self,gunnarsson1976exchange,perdew1992accurate}, to generalized gradient approximation (GGA)~\cite{perdew1996generalized,perdew2008restoring,becke1988density,lee1988development}, to meta-GGA (mGGA)~\cite{becke1989exchange,tao2003climbing,zhao2006new,ramasamy2025tackling,sun2012communication,del2012new, sun2015strongly,mardirossian2015mapping,tao2016accurate,mejia2017deorbitalization,furness2020accurate,lebeda2024balancing}, and to hybrid density functionals~\cite{becke1993new,Becke1993a,perdew1996rationale,heyd2003hybrid,xu2004x3lyp,zhao2008m06,csonka2010global,mardirossian2016omegab97m}. This evolution has largely been driven by the systematic incorporation of exact constraints, fundamental mathematical properties of the XC energy and its underlying XC hole density, that improve accuracy. For example, LSDA's success stems from satisfying exact XC hole constraints~\cite{langreth1975exchange, gunnarsson1976exchange}, while GGAs like the Perdew-Burke-Ernzerhof (PBE) functional~\cite{perdew1996generalized} and mGGAs like the strongly constrained and appropriately normed (SCAN) density functional~\cite{sun2015strongly} were developed to fulfill additional exact constraints. Other successful density functionals~\cite{zhao2006new, zhao2008m06, mardirossian2015mapping, Becke1993a, xu2004x3lyp, mardirossian2016omegab97m, kirkpatrick2021pushing} have been developed using total energy reference datasets.

Despite these advances, significant challenges such as self-interaction errors \cite{perdew2008density,haasler2020local, perdew1981self,pederson2014communication,li2018localized,zhang2020symmetry} and strong correlation \cite{becke2013density, su2018describing,johnson2011communication, perdew2021interpretations,mccarty2020bypassing} persist, often requiring nonlocal treatments.
However, aligning nonlocal corrections with semilocal DFAs, especially GGAs and mGGAs, remains difficult due to the gauge ambiguity of the XC energy density
\cite{burke1998unambiguous}. 
This mismatch complicates the development of nonlocal functionals, as seen in local hybrid density functionals~\cite{perdew2008density, arbuznikov2014towards, arbuznikov2024suppressing} and self-interaction corrections~\cite{perdew1981self,pederson2014communication,zope2019step}. {\color{red}All gauges are physically equivalent; the difficulty arises when a semilocal DFA exchange energy density is combined with a nonlocal one that is not expressed in the same gauge. Addressing these issues therefore requires constructing semilocal DFAs consistently within the gauge of the nonlocal ingredient, which for exact exchange is conventionally defined and calculated in the Hartree gauge (HG)~\cite{bhattarai2020step}.}
This approach not only facilitates the development of nonlocal density functionals through gauge alignment but also enables producing extensive reference XC energy density datasets for machine learning applications.

{\color{red}
There have been efforts that used features of exact exchange in HG to develop DFAs.
For example, Axel Becke's pioneering B88 \cite{becke1988density} exchange GGA recovers the correct asymptotic behavior of the exact exchange energy density in HG. It also recovers the second-order gradient expansion of the exchange energy with a coefficient approximately twice the first-principles value, resulting in poor performance for extended systems like jellium surfaces. 
The Becke–Roussel (BR89) exchange model~\cite{becke1989exchange} advanced this approach by constructing a mGGA from a hydrogen-like, spherically averaged exchange hole. Its parameters are fixed locally by the hole normalization and short-range curvature, incorporating the density Laplacian and kinetic energy density. Exact for hydrogen in its original form, the model hole places BR89 naturally within what we now call the HG, making it a conceptual predecessor of the present work. However, BR89 does not have an explicit closed form and an approximate self-consistent implementation was achieved only later by Proynov, Gan, and Kong \cite{proynov2008analytical}.
}


Here, we propose an exchange mGGA, called SORFKL, explicitly constructed in HG while maximally satisfying exact constraints of the exact exchange energy. SORFKL recovers the exact exchange HG energy density for the uniform electron gas (UEG) and closely matches it for the hydrogen atom, thus restoring the exact exchange HG energy density for the asymptotic and core regions of molecules and solids. The parameters of SORFKL are then fixed by appropriate norms, specifically the exchange energies of rare-gas atoms and jellium surfaces, following SCAN's design strategy.
Restoring the asymptotic and core regions of the exact exchange HG energy density and guided by exact constraints and appropriate norms, SORFKL provides an accurate representation of the exact exchange HG energy density for molecules at equilibrium with diverse bonding characteristics.

\begin{figure}
    \centering
    \includegraphics[width=1\linewidth]{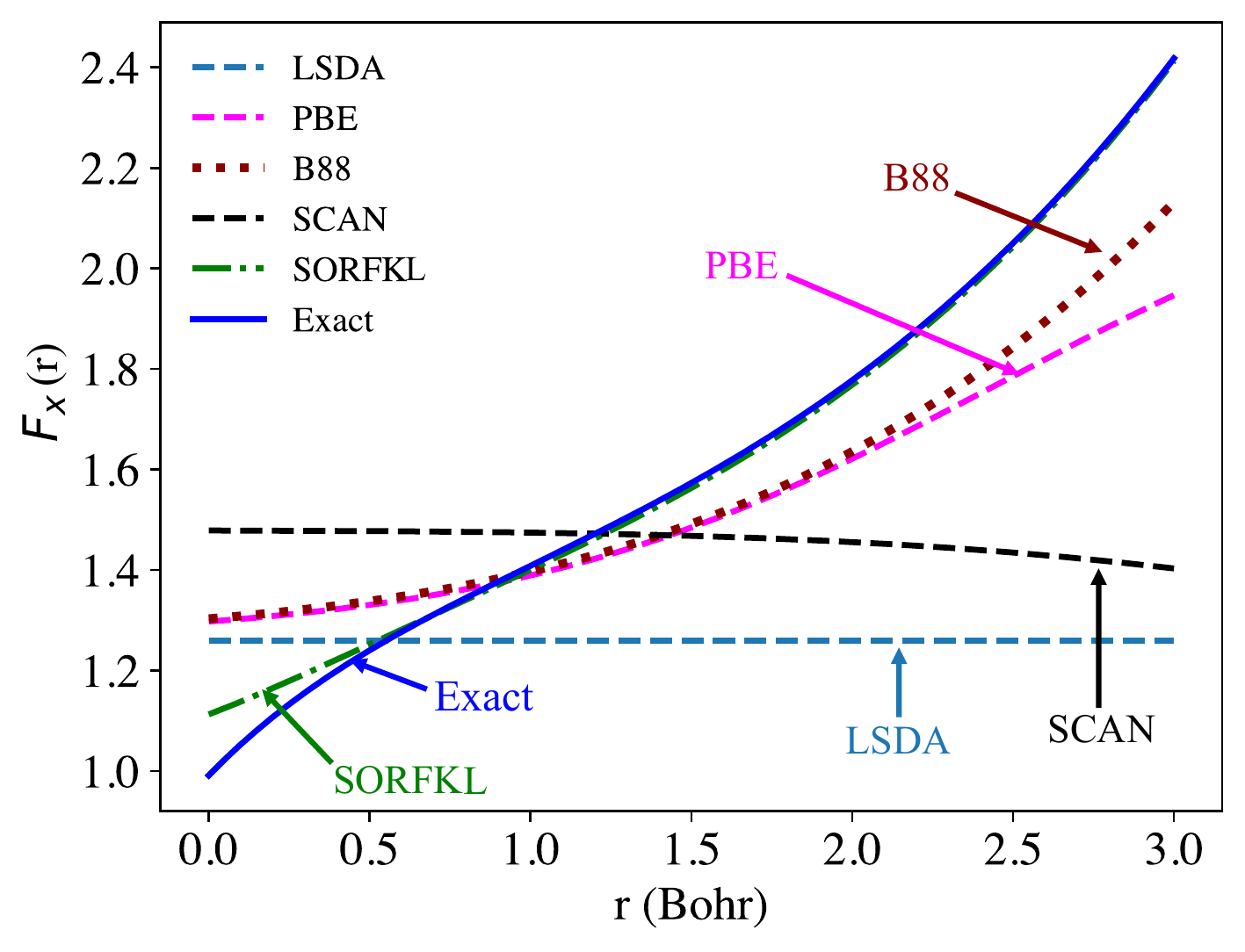}
    \caption{Exchange enhancement factor of the hydrogen atom for different density functionals including LSDA, PBE, B88, SCAN, and SORFKL, and the exact exchange. The analytical electron density of the hydrogen atom is used. The corresponding exchange energies in Hartree are $-0.2680$ (LSDA), $-0.3059$ (PBE), $-0.3098$ (B88), $-0.3125$ (SCAN), $-0.3125$ (SORFKL), and $-0.3125$ (Exact). 
    }
    \label{fig:Fx_H}
\end{figure}

The exact exchange energy as a functional of the electronic density $n(\br)$ is uniquely defined and can be expressed as 
\begin{equation}
    \label{eq:Exdef}
    E\x\exx[n] =-\frac{1}{2} \int d\br \int d\brP~\frac{~ |\rho_1(\br, \brP)|^2 }{|\br - \brP|}~,
\end{equation}
where $\rho_1(\br, \brP)=\Sigma_i^{occ}\phiiS \phiiP$ is the one-electron reduced density matrix defined by the occupied KS orbitals  $\left\{  \phi_i[n](r) \right\}$. In contrast, the  exchange energy density $e\x\exx(\br)$ is not unique. In fact, any nontrivial gauge transformation $e\x\exx(\br) \mapsto e\x\exx(\br) + G(\br)$ with $\int d\br G(\br) =0$ changes $e\x\exx(\br)$ but not $E\x\exx[n]$. {\color{red} A common choice, adopted throughout this work, is the HG,} in which $e\x\exx(\br)/n(\br)$ corresponds directly to half of the electrostatic potential from the exchange hole,
\begin{align}
    e\x\exx(\br)/n(\br) 
    &=  \frac{1}{2} \int d\brP \frac{n\x(\br, \brP)}{|\br - \brP|}~.
\end{align}
The exchange hole $n\x(\br, \brP)=-|\rho_1(\br, \brP)|^2 /n(\br)$ is a component of the XC hole, whose system average is uniquely defined and an observable \cite{perdew1996generalizedhole}.
Semilocal DFAs can produce accurate exchange energies for systems with localized exchange holes, whereas they typically yield too negative exchange energies for systems with a delocalized hole~\cite{sun2015strongly}. Thus, comparing a semilocal exchange energy density, if constructed in HG, with the exact exchange HG energy density can, in principle, reveal whether the underlying exact exchange hole is localized or delocalized, an essential insight for developing nonlocal density functionals~\cite{perdew2008density}. However, since the exchange energy densities of semilocal DFAs, $e\x\sll(\br)$, are usually not in the HG~\cite{tao2008exact}, a direct comparison between $e\x\sll(\br)$ and $e\x\exx(\br)$ is of limited significance. In contrast, for a semilocal DFA constructed in HG, such as SORFKL, the ratio $e\x\sll(\br) / e\x\exx(\br)$ can reliably indicate whether the underlying exact exchange hole is localized.


For a spin-unpolarized UEG, a cornerstone in almost all DFAs, the exchange energy density is given by $e\x^{\text{uni}}(\br) = -\frac{3}{4\pi} (3\pi^2)^{1/3} n(\br)^{4/3}$.
In HG, the enhancement factor of exact exchange is \(F\x\exx(\br) = e\x\exx(\br) / e\x^{\text{uni}}(\br)\), while for a semilocal DFA, typically not in HG, the exchange enhancement factor is \(F\x^{\text{sl}}(\br) = e\x^{\text{sl}}(\br) / e\x^{\text{uni}}(\br)\).

Upon assumption of local UEG behavior, the LSDA exchange DFA integrates \(e\x^{\text{uni}}(\br)\) over all space, followed by spin-scaling~\cite{oliver1979spin}. LSDA is in HG, as its energy density originates from the UEG exchange hole. However, due to its neglect of electron density inhomogeneity, LSDA is a poor approximation for inhomogeneous systems. Consequently, its enhancement factor significantly deviates from the exact exchange enhancement factor of the hydrogen atom (H) in HG, as shown in Fig.~\ref{fig:Fx_H}.

Alongside the UEG, the hydrogen atom serves as another paradigm system, with its electronic structure forming the foundation of much of chemistry. Specifically, the ground state of H, i.e., the 1s orbital, largely dictates the behavior of core electrons in molecules and solids. Accurately restoring the exact exchange HG energy density of H is therefore essential for developing exchange density functional approximations in HG for the core regions of both molecules and solids.  

As shown in Fig.~\ref{fig:Fx_H}, none of the widely used semilocal DFAs shown there recovers
the exact exchange HG energy density of H; BR89, discussed above, is a notable exception, being exact for H by construction. The B88 exchange GGA significantly improves over LSDA but still substantially deviates from the exact exchange HG enhancement factor. Meanwhile, the SCAN exchange mGGA~\cite{sun2015strongly}, although satisfying all six exact constraints suitable for semilocal exchange, deviates from the exact HG exchange enhancement factor of H even more than LSDA, indicating that SCAN is constructed in a gauge fundamentally different from the HG.

Remarkably, the SORFKL exchange enhancement factor closely matches the exact exchange HG enhancement factor of H, as shown in Fig.~\ref{fig:Fx_H}.
The SORFKL exchange energy for a spin-unpolarized electron density $n(\br)$ is defined as
\begin{equation}
    \label{eq:ExmGGA}
    E\x^\textrm{SORFKL}[n] = \int d\br~e\x^{\text{uni}}(n) F\x^\textrm{SORFKL}(s,\beta)~,
\end{equation}
where $F\x^\textrm{SORFKL}(s,\beta)$ is a
function of the reduced density gradient $s = |\nabla n| / (2(3\pi^2)^{1/3}n^{4/3})~$
and the iso-orbital indicator $\beta = (\tau-\tau\vW) / (\tau+\tau\uni)$ that identifies different chemical environments, with the range $0\leq \beta \leq 1$. The positive kinetic energy density of occupied KS orbitals is $\tau = \frac{1}{2}\sum_{i=1}^{\mathrm{occ}}|\nabla \phi_i|^2$, which reduces to $\tau\vW = {|\nabla n|^2}/{8n}$
in single-orbital systems ($\beta=0$)
and to $\tau\uni = ({3}/{10})(3\pi^2)^{2/3}n^{5/3}$
for a UEG ($\beta=1/2$). In noncovalent bonds, $\beta>> 1/2$ \cite{sun2013density,furness2019enhancing}. We choose $\beta$ over the related iso-orbital indicator $\alpha = (\tau-\tau\vW) / \tau\uni$ that can also identify different chemical environments~\cite{sun2013density, sun2015strongly, lebeda2025meta} due to $\beta$'s inherent numerical stability~\cite{furness2019enhancing}. The exchange energy of spin-polarized electron densities can be obtained by applying the exact spin-scaling relation \cite{oliver1979spin}.

$ F\x^\textrm{SORFKL}(s,\beta)$ is based on a model derived from the exact exchange HG energy density of the hydrogen atom,
\begin{equation}
    \label{eq:FxHyd}
    F\x\Hyd(s) = (3s)^{-2} 
    \Bigg[
    \frac{2}{3} \frac{ (s/s\nog)^3 - 1}{\ln\left( s/s\nog \right)} - 1
    \Bigg]~,
\end{equation}
where $s\nog=(6\pi)^{-1/3}$ is the smallest value of $s$ in H, occurring at the nucleus
(see the Supplemental Material for details \cite{supplemental,almbladh1985exact,march1987asymptotic,becke1988density}).
As $F\x\Hyd(s)$ can be negative for $s< s\nog$ and even diverges as $s\to0$, it is not suitable as a general model for exchange. Nevertheless, using $F\x\Hyd(s)$ as a framework provides several advantages.
First, $F\x\Hyd(s)$ naturally recovers the asymptotic behavior of the exact HG exchange energy density per particle $\varepsilon\x\exx=e\x\exx/n$ for finite systems, $\lim_{r\to\infty} \varepsilon\x\exx(\br) \sim -\frac{1}{2r}~$, by satisfying $\lim_{s\to \infty}F\x\Hyd(s) =(4\pi/9)\frac{s}{\ln(s)}~$, a condition also built into the B88 GGA. Furthermore, the core electrons of molecules and materials are dominated by the 1s orbitals that closely resemble the ground state 1s orbital of the hydrogen atom. Thus, designing a semilocal DFA that recovers $F\x\Hyd(s)$ for the core and asymptotic regions ensures gauge alignment with the exact exchange HG energy density in those regions.

To keep the desired features of $F\x\Hyd(s)$ and at the same time remove its unphysical behavior for $s<s_0$, we replace $s$ in Eq.~\eqref{eq:FxHyd} with a parameterizable function $g(s,\beta)$,
\begin{equation}
    F\x^\textrm{SORFKL}(s,\beta) = F\x\Hyd(g(s,\beta))~,
\end{equation}
where
\begin{equation}
\label{eq:g}
    g(s,\beta) = \left[ \frac{\sum_{i=0}^{N-1} a_i(\beta) s^{2i} + a_N s^{2N}}{\sum_{i=0}^{N-2} b_i(\beta) s^{2i} + a_N s^{2(N-1)}} \right]^{1/2}
\end{equation}
depends on $\beta$ through coefficients $\{a_i(\beta)\}$ and $\{b_i(\beta)\}$,
\begin{align}
    \label{eq:ai}
    a_i(\beta) &= \sum_{j=0}^M a_{ij}(1-2\beta)^j~\mathrm{for}~i=0\ldots N-1~,\\
    \label{eq:bi}
    b_i(\beta) &= \sum_{j=0}^M b_{ij}(1-2\beta)^j~\mathrm{for}~i=0\ldots N-2~.
\end{align}
We find that $N = 2$ and $M = 2$ provides sufficient flexibility for this work. Without loss of generality, $b_{00}=1$.

We demand $a_N$ to be independent of $\beta$ such that $\lim_{s\to \infty}g(s,\beta)=s$. Thus, $\lim_{s\to \infty}F\x^\textrm{SORFKL}(s,\beta)=\lim_{s\to \infty}F\x\Hyd(s)$ recovers the correct asymptotic behavior for finite systems. Additionally, this makes $F\x^\textrm{SORFKL}(s,\beta)$ only weakly dependent on $\beta$ for large $s$, a common feature of successful mGGAs \cite{tao2003climbing, sun2015strongly,task,lebeda2024balancing}.  
 
In iso-orbital regions ($\beta=0$), we demand $g(s,\beta=0) \simeq s$ for $s> s\nog$ such that $F\x^\textrm{SORFKL}(s,\beta=0)$ recovers $ F\x\Hyd(s)$ particularly for H.

At the center of two identical, infinitely separated closed shell atoms $s=0$, $\beta \to 1$, and $F\x^\mathrm{exact} \to \infty$, since this is also the asymptotic region of this system, with respect to the nuclei where electron densities concentrate.
To recover this condition we let the denominator of \Cref{eq:g} to zero for ($s=0$, $\beta = 1$) by (using $b_{00}=1$)
\begin{equation}
\label{eq:beta1}
    b_{01} = b_{02} + 1~.
\end{equation}


For the $H_2^+$ molecular ion, $\beta=0$ everywhere and, due to symmetry, $s=0$ at the bond center, regardless of the bond length $R$. In the limit $R \to 0^+$, where the two protons are maximally compressed, the electron density reduces to the hydrogenic $1s$ orbital density of He$^+$. The latter has the same exact exchange HG enhancement factor as H. This indicates that $F\x^\mathrm{exact}$ at the bond center of $H_2^+$ in the limit $R \to 0^+$ equals that of H at the nuclear center. We demonstrate this numerically in the Supplemental  \cite{supplemental,vydrov2009nonlocal,ramasamy2025tackling}. Consequently, we demand $F\x^\textrm{SORFKL}(s=0, \beta=0) = F\x^{Hyd}(s_0)$. Using Eq.~\eqref{eq:beta1} and $b_{00}=1$, this condition leads to
\begin{equation}
    a_{00} + a_{01} + a_{02} = 2 b_{01} s_0^2~.
    \label{fx_center}
\end{equation}

For the slowly varying density limit, i.e., $\beta(\br)\to 1/2$ and $s \to 0$, we recover the second-order gradient expansion for the exchange energy.
This fixes the coefficient $a_{00}= 0.2948$ and relates $a_{01}$, $a_{10}$, $a_2$, and $b_{01}$ by
\begin{equation}
    \label{eq:FxLim2nd}
    \frac{a_{2}}{b_{00}}=\frac{a_{10}}{a_{00}}-\frac{20}{81\,C_2}-\frac{65}{54} \left( \frac{b_{01}}{b_{00}}-\frac{a_{01}}{a_{00}}\right)~,
\end{equation}
with $C_2 =0.4380$, as detailed in the Supplemental Material \cite{supplemental,antoniewicz1985kohn,perdew1985accurate,gross1981gradient,brack1976extended, furness2022construction, sun2015semilocal}.

To ensure smoothness of $F\x^\textrm{SORFKL}(s,\beta=0)$ between $s=0$ and $s=s_0$, we further demand
\begin{equation}
\label{eq:Fx_sm_s}
    F\x^\textrm{SORFKL}(s,\beta=0) \geq F\x^{Hyd}(s\nog)~\text{for }0\leq s \leq s\nog~.
\end{equation}

Using Eqs.~\eqref{eq:beta1}--\eqref{eq:Fx_sm_s} and the exact exchange energy of H as constraints, we determine the coefficients $\{a_{ij}\}$ and $\{b_{ij}\}$
by fitting to the following appropriate norms: the exchange energies of rare-gas atoms and jellium surfaces \cite{sun2015strongly}.
In addition, we use the exact exchange HG energy density of H, the sole input directly utilized at the energy density level, to ensure gauge alignment. The final parameters, reference energies, and details on the optimization are summarized in the Supplemental Material \cite{supplemental,price2005differential, virtanen2020scipy,becke1988density, furness2022construction, pitarke2001jellium, furness2022construction,clementi1974roothaan,furness2021hartree,sun2015strongly,tao2008exact,jorge2000universal,smith2020psi4,lemmon2010nist,dunning1989gaussian, wilson1996gaussian}.

The resulting $F\x^\textrm{SORFKL}$ is smooth
and fulfills the conditions specified by Eqs.~\eqref{eq:beta1}--\eqref{eq:Fx_sm_s} (see Fig.~S2 of the Supplemental Material \cite{supplemental}).
Due to condition \eqref{eq:Fx_sm_s}, $F\x^\textrm{SORFKL}(s_0,\beta=0) > F\x^{Hyd}(s_0)$. Although this sacrifices some gauge accuracy around nuclei, where $F\x^\textrm{exact} \approx F\x^{Hyd}(s_0)$, we prioritize smoothness here, as the latter is an important design criterion for numerical stability~\cite{Medvedev2017,*Medvedev2017commentKepp,*Medvedev2017Response,furness2020accurate}.

The six exact constraints suitable for semilocal exchange DFAs are \cite{sun2015strongly,kaplan2023predictive}: (1) negativity, (2) spin-scaling, (3) uniform density scaling, (4) the gradient expansion for slowly varying electron densities, (5) non-uniform density scaling, and (6) the tight bound for single-orbital densities. In HG, we must sacrifice the constraints (5) and (6) that require \( \lim_{s \to \infty} F\x(s, \beta) \sim s^{-1/2} \) and \( F\x(s,\beta=0) \leq 1.174 \), respectively. These requirements contradict the asymptotic behavior of the exact exchange HG energy density, $\lim_{s \to \infty} F\x(s, \beta) \sim s/\ln(s)$. Constraints (5) and (6) are, together with the gradient expansion discussed next, the reason for SCAN's significant deviation from the HG.

In HG, the gradient expansion of the exchange energy density for slowly varying densities is not well-defined due to an oscillatory term involving the Laplacian of the electron density that is not integrable~\cite{gross1981gradient,armiento2002subsystem}. This non-integrable term can be removed via integration by parts, transforming the dependence on the Laplacian of the density into a dependence on the (reduced) density gradient~\cite{gross1981gradient,perdew1985accurate}. Although this procedure yields a well-defined gradient expansion for the exchange energy, the resulting exchange energy density is no longer in the HG~\cite{bhattarai2020step}. For a semilocal DFA, it is inherently impossible to match the nonlocal nature of the exact exchange HG energy density for slowly varying densities. However, given the importance of the slowly varying density limit for solids \cite{perdew2008restoring,sun2015strongly}, SORFKL is designed to satisfy the second-order gradient expansion of the exchange energy.

\begin{table}[htb]
\caption{Mean absolute percentage errors (MAPE) of exchange functionals for exchange energies of rare-gas atoms and jellium surfaces. See the Supplemental Material \cite{supplemental,becke1988density,furness2022construction,pitarke2001jellium} for detailed values.}
\begin{ruledtabular}
{\color{red}
\setlength{\tabcolsep}{1.25pt}
\small
\begin{tabular}{lcccccc}
                 &  LDA & PBEsol &  PBE &  B88 & SCAN & SORFKL \\ \hline \hline
\vspace{-0.3cm}\vspace{-\baselineskip}\\
rare gas atoms   & 6.75 &  3.02  & 0.48 & 0.12 & 0.29 & 0.32 \\ 
jellium surfaces & 45.5 &  2.75  & 21.1 & 28.8 & 24.5 & 0.62 \\
\end{tabular}
}
\end{ruledtabular}
\label{tab:1}
\end{table}

Table \ref{tab:1} demonstrates that SORFKL achieves outstanding accuracy for exchange energies of both rare-gas atoms and jellium surfaces, ranking among the best in each category. 
{\color{red} For comparison, B88, 
designed to recover the correct asymptotic behavior of the exchange energy density
and fitted to rare-gas exchange energies, achieves exceptional accuracy for rare-gas atoms but underperforms for jellium surfaces, whereas BR89 achieves a moderate mean absolute percentage error of 0.91 \% for rare-gas atom exchange energies~\cite{becke1989exchange}. }
In contrast, PBEsol, a revised version of PBE that was optimized for solids and recovering the first-principles second-order gradient expansion coefficient~\cite{perdew2008restoring}, achieves significantly improved accuracy for jellium surfaces, but at the cost of larger errors for rare-gas atoms.
This reflects the trade-offs in functional design for finite and extended systems at the GGA level.
Overall, SORFKL achieves accuracy for rare-gas atoms similar to the SCAN mGGA,
but is by far the most accurate for jellium surfaces, even surpassing PBEsol.

{\color{red}
As a first test beyond exchange, we combined SORFKL with PBE correlation~\cite{perdew1996generalized} in non-self-consistent calculations. Compared with exact exchange plus PBE correlation, this pairing improves atomization energies and lattice constants, but is less accurate for noncovalent interactions and barrier heights (see Sec.~IX of the Supplemental Material~\cite{supplemental}; see also Refs.~\cite{lynch2003small,jurecka2006benchmark,takatani2010basis,zheng2007representative,sun2011self,franzke2023turbomole,kresse1993ab,kresse1996efficiency,kresse1996efficient,birch1947finite,murnaghan1944compressibility} therein). A correlation functional designed to be compatible with the HG exchange is under development and will be reported elsewhere.
}

\begin{figure}[htb]
\centering
\includegraphics[width=\linewidth]{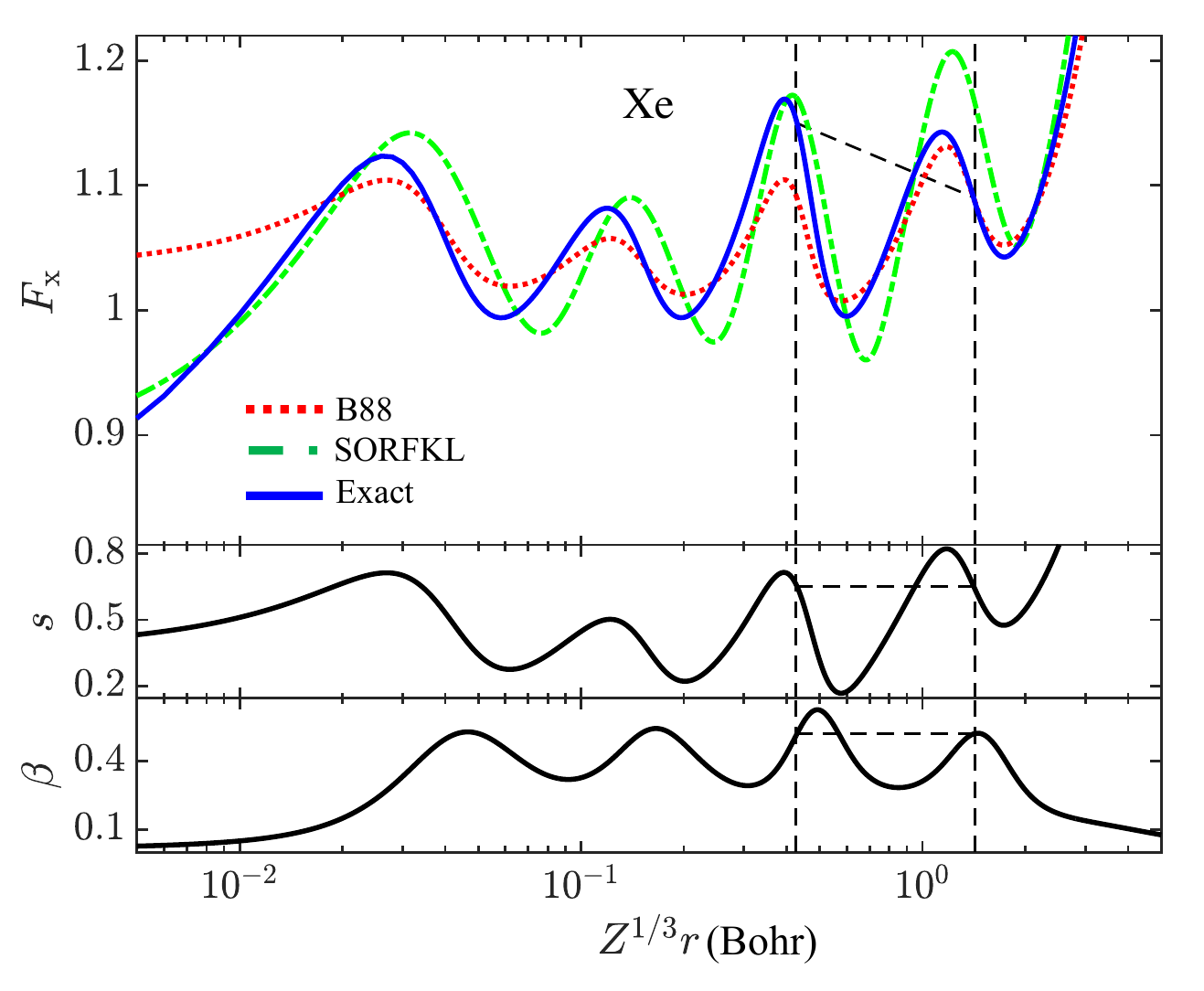}
\caption{Exchange enhancement factors of B88, SORFKL, and HG exact exchange, as well as $s$ and $\beta$ against $(Z^{1/3}r)$ for the xenon atom. Z is the nuclear charge.
}
\label{fig:3}
\end{figure}

Figure \ref{fig:3} compares the enhancement factors \(F\x\) of B88, SORFKL, and the exact exchange in HG as a function of \((Z^{1/3}r)\) for the xenon atom. Both B88 and SORFKL closely match $F\x^\mathrm{exact}$ in the tail region due to their recovery of the correct asymptotic behavior. Although B88 is highly accurate for the exchange energies of rare-gas atoms, it overestimates $F\x^\mathrm{exact}$ in the core region and in the shell regions where \(s\) is small. To compensate for this, B88 underestimates $F\x^\mathrm{exact}$ in the intershell regions where \(s\) is larger, resulting in damped oscillations compared to $F\x^\mathrm{exact}$. SORFKL, on the other hand, achieves similar accuracy for the exchange energies but improves over B88 especially in the core region (\(r=0\)). As a result, SORFKL aligns more closely with the exact exchange HG enhancement factor throughout all spatial regions.

\begin{figure}[htb]
\centering
\includegraphics[width=\linewidth]
{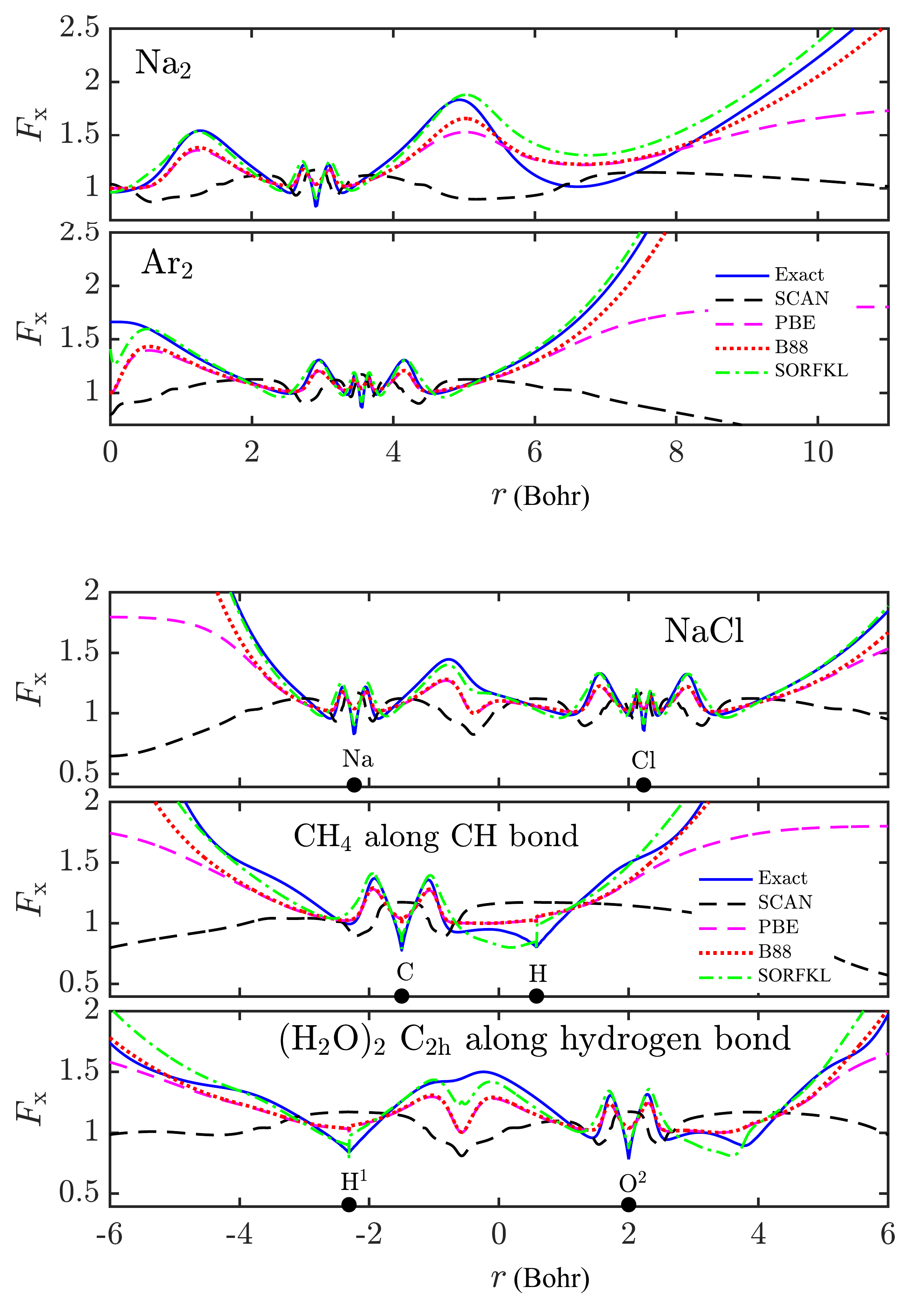}
\caption{
Exchange enhancement factors $F\x$ of selected DFAs and HG exact exchange along the bond axis of molecules representative of different chemical bond types. For Na$_2$ and Ar$_2$, the bond center is at $r=0$.
}
\label{fig:4}
\end{figure}

Despite these achievements, we should point out that it is impossible to reproduce $F\x^\mathrm{exact}$ exactly over the entire spatial range with a semilocal DFA. For example, the vertical dashed lines in Fig.~\ref{fig:3} correspond to two radial points with identical $(s,\beta)$ values but different $F\x^\mathrm{exact}$ of approximately $1.15$ and $1.09$, respectively. Such behavior is impossible to capture with any semilocal DFA.


Next, we assess SORFKL across a diverse set of molecules, each representing a distinct chemical bonding environment: CH$_4$ for covalent bonding, Na$_2$ for metal-metal bonding, NaCl for ionic bonding, Ar$_2$ for van der Waals interactions, and (H$_2$O)$_2$ for hydrogen bonding. 

Figure \ref{fig:4} once again highlights that SCAN is in a gauge very different from HG, as it exhibits significant deviations from the exact exchange HG enhancement factor $F\x^\mathrm{exact}$. Similarly to its performance on rare-gas atoms, B88 closely matches $F\x^\mathrm{exact}$ in the tail regions, but overestimates $F\x^\mathrm{exact}$ in the core and shell regions and underestimates $F\x^\mathrm{exact}$ in the intershell regions. Interestingly, for the bonding regions of the considered molecules, B88 generally underestimates $F\x^\mathrm{exact}$, except for the C--H bonds in CH$_4$.
PBE closely follows B88, except for the tail regions, where PBE does not recover the asymptotic behavior of the exact exchange HG energy density.

SORFKL outperforms B88 and PBE overall, achieving superior accuracy in the core, shell, intershell, and tail regions. Remarkably, SORFKL also closely tracks $F\x^\mathrm{exact}$ in the bonding regions, marking a significant improvement over B88.
These results indicate that SORFKL accurately captures the exact exchange HG energy density for systems in equilibrium, where the exchange holes are localized.

{\color{red}In conclusion, we developed SORFKL, a closed-form mGGA for exchange constructed in HG. SORFKL recovers the exact UEG exchange energy density, closely reproduces the exact exchange HG energy density of the hydrogen atom, thereby capturing the asymptotic and core regions in molecules and solids, and satisfies the second-order gradient expansion of the integrated exchange energy.} Following SCAN’s design strategy, it satisfies all exact exchange constraints compatible with HG and uses appropriate norms. Beyond its immediate performance, SORFKL provides a framework for nonempirical functional development by leveraging gauge alignment as a core principle, enabling more consistent nonlocal corrections and unlocking the potential of large reference XC energy density datasets for machine learning. This foundation positions SORFKL as a promising step toward next-generation density functionals that combine exact constraint satisfaction, gauge consistency, and data-driven design, with potential as a general-purpose mGGA when paired with a suitable correlation functional.

\section*{Data availability}
The data that support the findings of this article are available within the article and its Supplemental Material.


\section{Acknowledgments}

This work acknowledges the support of the U.S. DOE, Office of Science, Basic Energy Sciences Grant No. DE-SC0019350. T.L. acknowledges the support of National Science Foundation
(NSF) under Grant No. DMR-2042618. This work also used Purdue Anvil CPU at Rosen Center for Advanced Computing (RCAC) through allocation DMR190076 from the Advanced Cyberinfrastructure Coordination Ecosystem: Services \& Support (ACCESS) program, which is supported by National Science Foundation grants \#2138259, \#2138286, \#2138307, \#2137603, and \#2138296. We thank John Perdew for discussions. {\color{red} We dedicate this work to the memory of Axel D. Becke, whose pioneering contributions to density functional theory, particularly exchange-hole modeling and exchange-correlation functional development, laid important foundations for the present work.}


\bibliography{bib}

\end{document}


\title{Supplementary material for ``Meta-generalized gradient
approximation made in the Hartree gauge''}%
\author{Yan Oueis}
\thanks{These authors contributed equally to this work}%
\author{Akilan Ramasamy}%
\thanks{These authors contributed equally to this work}%
\author{James W. Furness}%
\altaffiliation[Current address: ]{Schr\"odinger, Inc., New York, New York 10036, USA}%
\author{Jamin Kidd}%
\author{Timo Lebeda}%
\author{Jianwei Sun}%
 \email[Corresponding Author:]{jsun@tulane.edu}
\affiliation{%
Department of Physics and Engineering Physics, Tulane University, New Orleans, LA 70118, USA
}%

\date{\today} 

\setcounter{figure}{0}
\renewcommand{\figurename}{Figure}
\renewcommand{\thefigure}{S\arabic{figure}}

\setcounter{equation}{0}
\renewcommand{\theequation}{S\arabic{equation}}
\renewcommand{\thetable}{S\arabic{table}}

\newcommand{\x}{_\mathrm{x}}
\newcommand{\mc}{_\mathrm{c}}
\newcommand{\Hyd}{^\mathrm{Hyd}}

\maketitle
\setcounter{secnumdepth}{3} 


Here, we provide figures of $g(s,\beta)$ and $F\x(s,\beta)$, a derivation of $F\x\Hyd$ and its asymptotic behavior, details on the second-order gradient expansion of $F\x^\mathrm{SORFKL}$, on SORFKL's parameters, and on the exchange energies of rare-gas atoms and jellium surfaces, and the computational details. Furthermore, we show enhancement factor plots for more rare-gas atoms and discuss the enhancement factor in the context of the $H_2^+$ binding energy curve. Finally, we report benchmarks of SORFKL combined with PBE correlation for atomization energies (AE6), noncovalent interactions (S22), barrier heights, and lattice constants (LC20), together with the corresponding computational details.

\section{Replacing $s$ with $g(s,\beta)$ and a comparison of $F\x^\mathrm{SORFKL}(s,\beta)$ with $F\x\Hyd(s)$, $F\x^\mathrm{B88}(s)$, and $F\x^\mathrm{PBE}(s)$}
\begin{figure}[ht!]
\centering
\includegraphics[width=\textwidth]{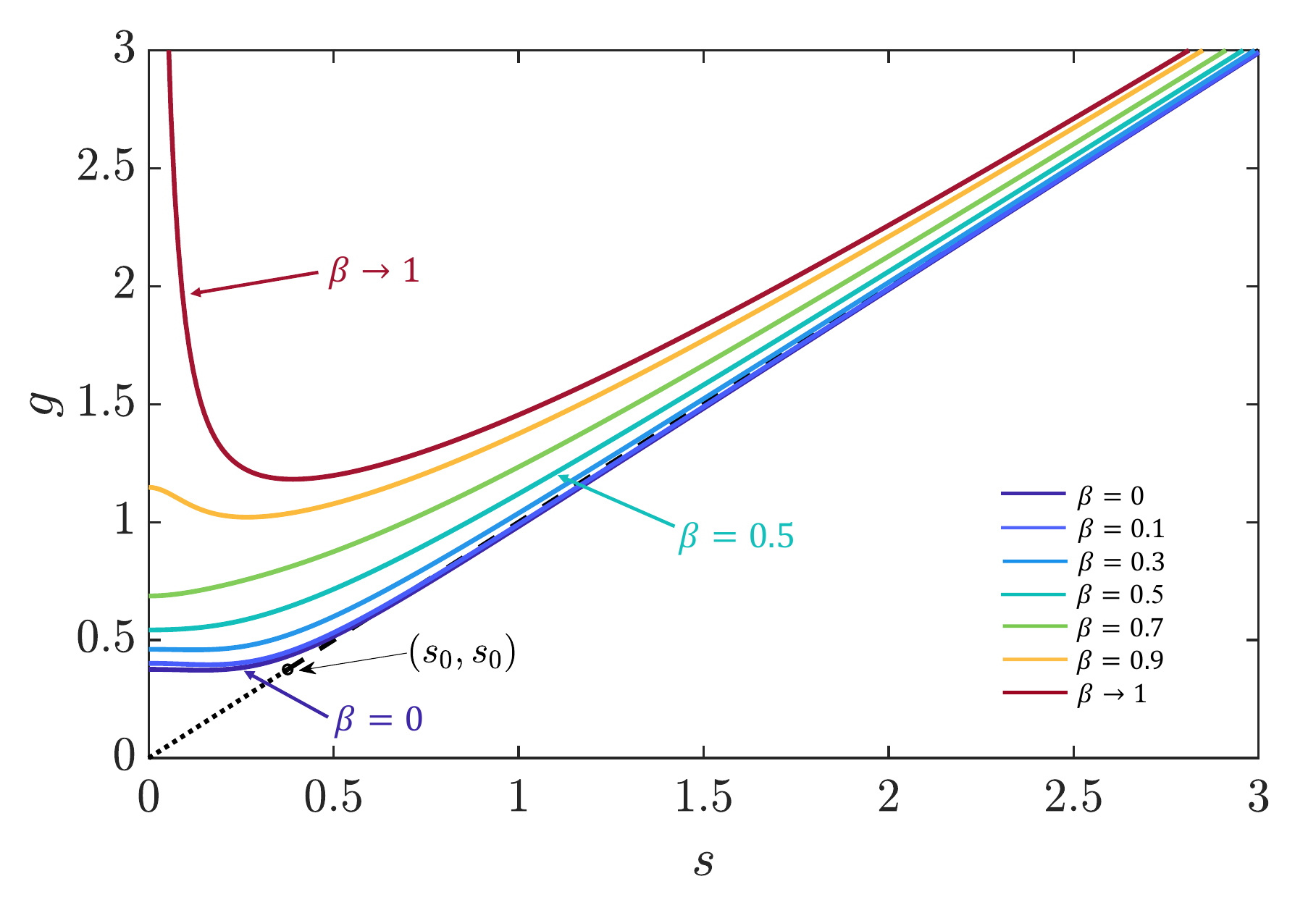}
\caption{$g(s,\beta)$ is plotted against s at different $\beta$. The line $g=s$ is plotted with a dashed line for $s\geq s_0$ and with a dotted line for $s\leq s_0$, where $s_0=(6\pi)^{-1/3}$ is the $s$ value at the hydrogen nucleus.}
\label{fig:1}
\end{figure}
\begin{figure}
\centering
\includegraphics[width=\textwidth]{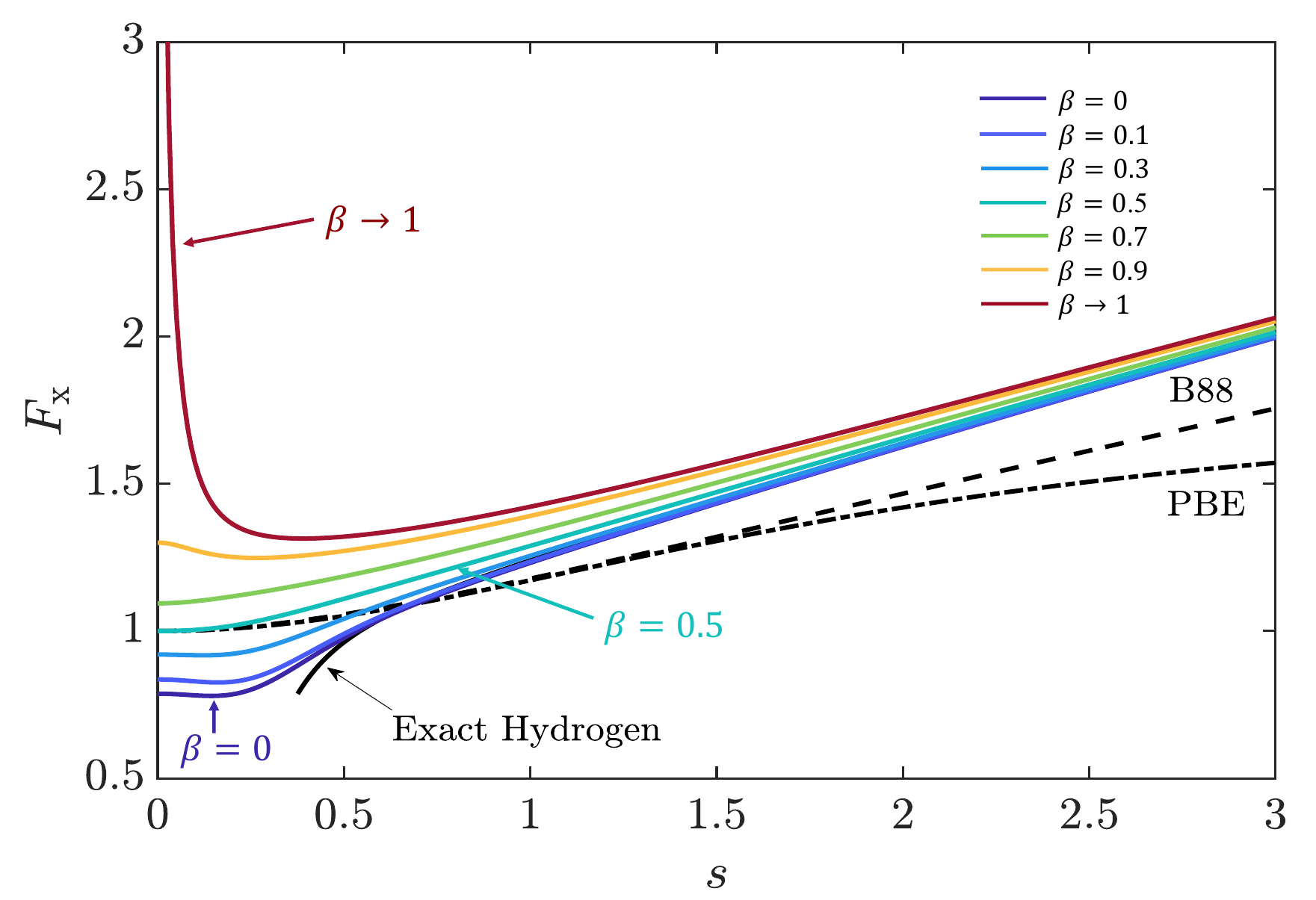}
\caption{$F\x^\textrm{SORFKL}(s,\beta)$ as function of s for different $\beta$ along with $F\x\Hyd(s)$ (solid), $F\x^\mathrm{B88}(s)$ (dashed), and $F\x^\mathrm{PBE}(s)$ (dot-dashed). $F\x\Hyd(s)$ is shown only for $s > s_0$ with $s_0=(6\pi)^{-1/3}$.}
\label{fig:2}
\end{figure}

\section{Derivation of $F\x^{Hyd}(s)$ and its asymptotic behavior}
\label{sec: FxHyd}





The fully spin-polarized electron density of the hydrogen atom is 
\begin{equation}
    n^H(r) = \frac{e^{-2r}}{\pi}~.
    \label{nh}
\end{equation}


For one-electron systems, the correlation energy vanishes and the exchange energy equals the negative of the Hartree energy. Consequently, the exchange energy can be expressed as
\begin{equation}
    E\x[n] = -\frac{1}{2} \int d^3r \int d^3r' \frac{n(\textbf{r})n(\textbf{r}')}{|\textbf{r}-\textbf{r}'|}~,
\end{equation}
and the exchange hole here is $n\x(\textbf{r}, \textbf{r}') = -n(\textbf{r}')$~.
Therefore, the exchange energy density per electron in the Hartree gauge (HG) is (after integrating over the spherical degrees of freedom)
\begin{equation}
    \varepsilon\x(r) =- \frac{1}{2}\Bigg[\int^{r}_{0} dr' \frac{4\pi r'^2 n(r')}{r} 
    + \int^{\infty}_{r} 4\pi r' n(r') \Bigg]~.
\end{equation}
Inserting the hydrogen electron density, Eq.~\eqref{nh}, the exact exchange HG energy density per electron of H reads
\begin{equation}
    \varepsilon^{exact}\x(r) =- \frac{1}{2r}\Big(1-e^{-2r}(1+r)  \Big)~,
    \label{exactex}
\end{equation}
Thus,
\begin{equation}
\begin{aligned}    
    F\x^{exact}(r) 
    &= \frac{\varepsilon\x^{exact}(r) }{\varepsilon\x^{uni}(r)} \\
    &= 2\left(\frac{\pi}{9}\right)^{2/3} \frac{e^{\frac{-4r}{3}}}{r} \Bigg[e^{2r} -r -1
    \Bigg]~,
    \label{fxhr_unpolar}
\end{aligned}
\end{equation}
where $\varepsilon\x^{uni}(n) = -3/(4\pi) (3 \pi^2 n)^{1/3}$.
Obviously, Eq.~\eqref{exactex} leads to $\lim_{r\to \infty} \varepsilon\x^{exact}(r)=-\frac{1}{2r}$.






The $F\x^{Hyd}(s)$ model is defined for spin-unpolarized systems. Therefore, after applying the spin scaling relationship, the exchange energy of the hydrogen atom calculated with the $F\x^{Hyd}(s)$ model is
\begin{equation}
    E\x^{Hyd}[n^H,0]= \frac{1}{2} \int d^3r e\x^{uni}(2n^H) F\x^{Hyd}(s[2n^H])~.
    \label{ex-Hyd}
\end{equation}
By requiring the exchange energy density of Eq.~\eqref{ex-Hyd} reproducing the exact exchange HG energy density, we have,
\begin{equation}
    F\x^{exact}(r) = 2^{1/3} F\x^{Hyd}(s[2n^H])~.
    \label{fx_exact_h}
\end{equation}
From the definition of $s$, we have 
\begin{equation}
    s[2n^H] = \frac{e^\frac{2r}{3}}{(6\pi)^{\frac{1}{3}}}~.
    \label{sh}
\end{equation}
By combining Eqs.~\eqref{fxhr_unpolar} and \eqref{fx_exact_h}, replacing $r$ with $s$, and rearranging terms, we arrive at Eq.~(4) of the main text.

For a finite system, the electron density at the asymptotic region decays exponentially as,
\begin{equation}
    \lim_{r\to \infty}n(\textbf{r}) = e^{-\delta r},
    \label{asymp_n}
\end{equation}
where $n^H(r)$ follows the same form. Here, $\delta$ is a constant related to the ionization potential of the system \cite{almbladh1985exact}. The corresponding reduced density gradient there diverges as $\lim_{r\to \infty} s= \frac{\delta e^{\delta r/3}}{2(3\pi^2)^{1/3}}$. Since $\lim_{s \to \infty} F\x^{Hyd}(s)=\frac{4\pi}{9} \frac{s}{ln(s)}$, any enhancement factor $F\x^{mGGA}$ recovering $\lim_{s \to \infty}F\x^{Hyd}(s)$ leads to $\lim_{r\to \infty} \varepsilon\x^{mGGA}(r)=-\frac{1}{2r}$, which is the exact asymptotic behavior of the exchange energy density per electron of the finite system in the Hartree gauge \cite{march1987asymptotic}. These have been carefully discussed in Ref.~\cite{becke1988density}.

\section{The second-order gradient expansion of $F\x^{SORFKL}(s, \beta)$}

\label{sec_deriv}

The second-order gradient expansion for the exchange energy of a slowly varying electron density can be satisfied by requiring \cite{antoniewicz1985kohn,perdew1985accurate,gross1981gradient}
\begin{equation}
\label{eq:FxLim}
     \lim_{|\nabla n| \to 0} F\x^\textrm{SORFKL}(s,\beta) = 1 + \frac{10}{81}s^2~.
\end{equation}

Using the gradient expansion of $\tau$~\cite{brack1976extended},
\begin{equation}
    \tau = \tau^{uni}\Bigg[1 + \frac{20}{9}q + \frac{5}{27}s^2 \Bigg] +\mathcal{O}[|\nabla n|^4]~,
\end{equation}
where $q = \nabla^2n/[4(3\pi^2)^{2/3}n^{5/3}]$ is the reduced electron density Laplacian, we have
\begin{equation}
    \lim_{|\nabla n|\rightarrow 0} (1-2\beta) = -\frac{10}{9}q + \frac{85}{54}s^2 + \mathcal{O}(|\nabla n|^4)~.
\end{equation}
Under integration by parts~\cite{furness2022construction, sun2015semilocal}, $q$ can be recast in terms of $s^2$, leading to
\begin{equation}
    q \overset{p.I.}{\cong} s^2/3~.
\end{equation}
The second order gradient expansion of $g(s, \beta)$ then reads
\begin{equation}
    \lim_{|\nabla n|\rightarrow 0} g(s, \beta) = \Big(\frac{a_{00}}{b_{00}}\Big)^{\frac{1}{2}}(1 + \Delta g^{(2)} (s))~,
\end{equation}
with
\begin{equation}
    \Delta g^{(2)}(s) = \Bigg[\frac{65}{108} \Big(\frac{a_{01}}{a_{00}} - \frac{b_{01}}{b_{00}}\Big) + \frac{1}{2}\Big(\frac{a_{10}}{a_{00}} - \frac{a_{2}}{b_{00}}\Big)\Bigg]s^2~.
    \label{delta_g}
\end{equation}

Thus, 
\begin{equation}
\begin{aligned}
    \lim_{|\nabla n| \rightarrow 0} F\x^\textrm{SORFKL}(s, \beta)=C_1+C_2\Delta g^{(2)}(s)~,  
\end{aligned}
\end{equation}
where 
\begin{equation}
    C_1=\frac{1}{9} \Bigg[
    \frac{\frac{2}{3} 
    \left( 
    6\pi \sqrt{\frac{a_{00}}{b_{00}}} - \frac{b_{00}}{a_{00}} 
    \right)}{
    (\frac{1}{3}) \ln(6\pi) + \frac{1}{2} \ln \Big(\frac{a_{00}}{b_{00}}\Big)
    } - \frac{b_{00}}{a_{00}} \Bigg]~,
    \label{ueg}
\end{equation}
and 

\begin{equation}
\begin{aligned}
    C_2 &= \frac{2}{27} \Bigg[ 
    \frac{6\pi \sqrt{\frac{a_{00}}{b_{00}}} + \frac{2 b_{00}}{a_{00}}}
    {\ln{(6\pi)^{1/3}} + \frac{1}{2}\ln{\frac{a_{00}}{b_{00}}}} 
    - 
    \frac{6\pi \sqrt{\frac{a_{00}}{b_{00}}} - \frac{b_{00}}{a_{00}}}
    {\big(\ln{(6\pi)^{1/3}} + \frac{1}{2}\ln{\frac{a_{00}}{b_{00}}}\big)^2}
    \Bigg] + \frac{2}{9} \frac{b_{00}}{a_{00}}~.
\end{aligned}
\end{equation}

To recover the second order gradient expansion, we require 
$C_1=1$, leading to
\begin{equation}
    \frac{a_{00}}{b_{00}} \approx 0.29480~,
    \label{a_00}
\end{equation}
(note that we set $b_{00}=1$ without loss of generality and thus $a_{00}\approx 0.2948$) and $C_2 \Delta g^{(2)}(s)=\frac{10}{81}s^2$, leading to
\begin{equation}
    \label{eq:FxLim2}
    \frac{a_{2}}{b_{00}}=\frac{a_{10}}{a_{00}}-\frac{20}{81\,C_2}-\frac{65}{54} \left( \frac{b_{01}}{b_{00}}-\frac{a_{01}}{a_{00}}\right)~,
\end{equation}
which relates $a_{10}$, $a_{01}$, $a_2$, and $b_{01}$.

\section{SORFKL's parameters}
\label{sec:param}


Table \ref{tab:1} summarizes the parameters of SORFKL. For the optimization of SORFKL's parameters, we used the differential evolution algorithm \cite{price2005differential} as implemented in SciPy.Optimize library version 1.11.1 \cite{virtanen2020scipy}. 

\begin{table}[htb]
\caption{Values of parameters for $F\x^\textrm{SORFKL}(s,\beta)$.  
}
\begin{ruledtabular}
\begin{tabular}{ccccc}
$a_{00}$   & $a_{01}$   & $a_{02}$   & $a_{10}$   & $a_{11}$   \\ 
$0.2948$   & $0$  & $-0.0253$   & $3.9226$    & $-5.6680$   \\ \hline \hline
$a_{12}$   & $b_{00}$   & $b_{01}$   & $b_{02}$   & $a_{2}$    \\ 
$2.8873$    &  $1$       & $0.9545$   & $-0.0455$  & $11.5935$    \\ 
\end{tabular}
\end{ruledtabular}
\label{tab:1}
\end{table}

The total loss function was constructed as a weighted sum of scaled loss functions for the exchange energies of the rare-gas atoms, the jellium surface exchange energies, and the exact exchange HG energy density of the hydrogen atom (the latter as measure of hydrogen atom gauge accuracy).

The scaled loss functions were computed using the following general equation:
\begin{equation}
\label{eq:scaledloss}
     \tilde{\mathcal{L}}=\frac{\mathcal{L}}{\epsilon_\mathrm{max}}~,
\end{equation}
where $\mathcal{L}$ and $\epsilon_\mathrm{max}$ are the loss functions and the maximum reference errors respectively. 
For the rare-gas atoms, $\mathcal{L}$ is the MAPE in $E\x$ for Ne, Ar, Kr, and Xe,
and $\epsilon_\mathrm{max}$ is PBEsol's MAPE for the same atoms. For the jellium surface loss function, $\mathcal{L}$ is the MAPE in jellium surface exchange energies $\sigma\x$ for $r_s=2$, $r_s=3$, $r_s=4$, and $r_s=6$, and $\epsilon_\mathrm{max}$ is PBE's MAPE in $\sigma\x$ for the same bulk constants. 

The hydrogen atom gauge loss is defined as,
\begin{equation}
\label{eq:losshyd}
    \mathcal{L}^{Hyd} = \sum_{r_i=r_{min}}^{r_{max}} \left| e\x^{exact}(r_i) - e\x^\textrm{SORFKL}(r_i) \right|^2~,
\end{equation}
where the sum is computed over 1000 discrete radial points, ranging from \( r_{\text{min}} = 0.01 \) Bohr to \( r_{\text{max}} = 10 \) Bohr, covering both the near-nucleus and asymptotic regions.
The analytic electron density of hydrogen atom is used here. For the scaled $\tilde{\mathcal{L}}^\mathrm{Hyd}$, $\epsilon_\mathrm{max}$ is set to $5\times 10^{-3}$, which is arbitrary but a reasonably small number for the optimization.

Using the above definitions, the total loss functions can be written as:
\begin{equation}
    \tilde{\mathcal{L}}^\mathrm{Total}=\omega_1\tilde{\mathcal{L}}^\text{rare-gas}+\omega_2\tilde{\mathcal{L}}^\mathrm{Jellium}+\omega_3\tilde{\mathcal{L}}^\mathrm{Hyd}~,
\end{equation}
with hyperparameters $\omega_i$ responsible for relative contribution of individual scaled loss functions. Results presented in this paper correspond to 
$\omega_1=0.75$, $\omega_2=0.05$, and $\omega_3=0.2$. 
We observed, however, that the results of the optimization are not markedly affected by the values assigned to $\omega_i$'s as long as the relative contributions from the rare-gas atoms and jellium surfaces stay above $70\%$ and below $10\%$, respectively.

\section{Computational details}
\label{sec:comp}

The exchange energies of different density functionals and the exchange enhancement factor plots for rare-gas atoms were calculated from HF reference datasets with Slater-type orbitals~\cite{clementi1974roothaan,furness2021hartree}, with the reference HF exchange energies from Ref.~\cite{becke1988density}. For the jellium surfaces, LDA orbitals were used~\cite{sun2015strongly}.


Following Ref.~\cite{tao2008exact}, the Universal Gaussian Basis Set (UGBS) \cite{jorge2000universal} was used for calculating the enhancement factors of molecules shown in Figure 3 of the main text and for Figure ~\ref{fig:exact_fx_h2+}. The corresponding calculations were carried out in Psi4 \cite{smith2020psi4}. HF calculations provided density matrices that were used to derive HF electron densities and kinetic energy densities. These densities were then used to compute exchange enhancement factors of various functionals. Experimental molecular geometries were sourced from the NIST database \cite{lemmon2010nist} and used without further optimization.

The binding energy curves of $H_2^+$ were obtained using Psi4 with the cc-pV5Z basis set~\cite{dunning1989gaussian, wilson1996gaussian}.
The HF density was used in a non-self-consistent manner to evaluate the exchange energies for the considered density functionals. The binding energy curves for $H_2^+$ were generated based solely on the exchange functional, excluding the correlation functional contributions (which would be non-vanishing for the PBE GGA).

{\color{red}
For the benchmarks of SORFKL+PBEc and HF+PBEc in Sec.~\ref{sec:pbec} (Tables~\ref{tab:ae6}--\ref{tab:lc20}), the AE6, S22, and barrier-height calculations were carried out non-self-consistently on PBE0{\color{cyan}~\cite{perdew1996rationale}} densities with TURBOMOLE~\cite{franzke2023turbomole}, in which we implemented SORFKL directly in the source code so that SORFKL exchange could be combined with PBE correlation. The 6-311++G(3df,3pd) basis set was used for the AE6 and BH6 sets and
the aug-cc-pVTZ basis set for the S22 set using a fine integration grid (grid size 7). The LC20 calculations were performed with \textsc{vasp}{\color{cyan}~\cite{kresse1993ab,kresse1996efficiency,kresse1996efficient}}, non-self-consistently on fixed r$^2$SCAN orbitals and densities, which were regenerated self-consistently at each volume. We used a $\Gamma$-centered $16\times16\times16$ $k$-point mesh, \texttt{PREC = Accurate}, \texttt{EDIFF} $=10^{-5}$~eV, and a plane-wave cutoff converged for each system (e.g., 650~eV for Ag), with \texttt{ENAUG} scaled accordingly. Metals were treated with first-order Methfessel--Paxton smearing and semiconductors and insulators with Gaussian smearing. On the fixed orbitals, total energies were evaluated in a single electronic step (\texttt{ALGO = EIGENVAL}, \texttt{NELM = 1}), with PBE correlation combined with either Hartree--Fock exchange (\texttt{LHFCALC}, \texttt{AEXX = 1.0}; because \textsc{vasp} switches off correlation by default for \texttt{AEXX = 1.0}, PBE correlation was included explicitly through \texttt{ALDAC = 1.0} and \texttt{AGGAC = 1.0}) or SORFKL exchange. Equilibrium lattice constants were obtained by fitting the total energies computed at a series of volumes around equilibrium to the Birch--Murnaghan equation of state~\cite{birch1947finite,murnaghan1944compressibility}.
Reference values were taken from Refs.~\cite{lynch2003small} (AE6), \cite{takatani2010basis} (S22, using the CBS[$\Delta$a(DT)Z] interaction energies for the set of Ref.~\cite{jurecka2006benchmark}), \cite{zheng2007representative} (BH6), and \cite{sun2011self} (LC20).
}

%


\clearpage

\section{Exchange energies of rare-gas atoms and jellium surface for different $r_s$}

Tables \ref{tab:2} and \ref{tab:3} report the detailed data for the exchange energies of rare-gas atoms and jellium surfaces that supplement Table I of the main text.

\begin{table}[bht]
\caption{Exchange energies (Ha) of rare-gas atoms and the corresponding mean absolute percentage errors (MAPE) of the SORFKL and other functionals.}
\color{red}
\begin{ruledtabular}
\begin{tabular}{lccccc}
                                & Ne          & Ar          & Kr          & Xe          & MAPE (\%)   \\ \hline \hline 
Exact\footnotemark[1]           & $-12.108$   & $-30.188$   & $-93.89$    & $-179.2$    &    --   \\ \hline 
LDA                             & $-11.033$   & $-27.863$   & $-88.62$    & $-170.6$    & $6.75$   \\ 
PBEsol                          & $-11.665$   & $-29.143$   & $-91.45$    & $-175.0$    & $3.02$   \\
PBE                             & $-12.067$   & $-29.996$   & $-93.43$    & $-178.4$    & $0.48$   \\
B88                             & $-12.138$   & $-30.153$   & $-93.87$    & $-179.0$    & $0.12$   \\ 
BR89\footnotemark[2]            & $-12.19$    & $-30.09$    & $-92.88$    & $-176.4$    & $0.91$   \\ 
SCAN\footnotemark[3]            & $-12.164$       & $-30.264$       & $-94.071$       & $-179.315$       & $0.29$ \\
SORFKL    & $-12.203$   & $-30.204$   & $-93.774$    & $-178.639$    & $0.32$ 
\end{tabular}
\end{ruledtabular}
\footnotetext[1]{From \cite{becke1988density}}
\footnotetext[2]{From \cite{becke1989exchange}, with $\gamma=1$.}
\footnotetext[3]{From \cite{furness2022construction}}
\label{tab:2}
\end{table}
\begin{table}[bht]
\caption{Relative signed errors of the SORFKL and other functionals for the jellium surface exchange energies $\sigma\x$ (erg/cm$^2$) at various bulk density parameters $r_s$. The relative signed error is defined as $\frac{\sigma\x^{DFA}-\sigma\x^{exact}}{\sigma\x^{exact}}*100$
}
\begin{ruledtabular}
\begin{tabular}{lcccccc}
                                & $r_s=2$          & $r_s=3$          & $r_s=4$          & $r_s=6$          & MAPE (\%)    \\ \hline \hline 
Exact\footnotemark[1]           & $2624$      & $526$       & $157$       & $22$        &  --    \\ \hline 
LDA                             & $15.72$          & $27.15$          & $41.78$          & $97.27$          & $45.5$ \\ 
PBEsol                          & $1.589$          & $2.627$          & $3.295$          & $3.478$          & $2.75$\\   
PBE                             & $-7.150$         & $-11.73$         & $-18.45$         & $-47.08$         & $21.1$     \\  
B88                             & $-9.364$         & $-15.71$         & $-25.05$         & $-64.93$         & $28.8$ \\ 
SCAN\footnotemark[2]            & $0.27$            & $-7.02$           & $-19.18$          & $-71.53$         & $24.5$  \\
SORFKL    & $-1.527$         & $-0.760$         & $-0.197$         & $0.008$         & $0.62$  
\end{tabular}
\end{ruledtabular}
\footnotetext[1]{From \cite{pitarke2001jellium}}
\footnotetext[2]{From \cite{furness2022construction}}
\label{tab:3}
\end{table}

\section{Enhancement factor plots for more rare-gas atoms}

\begin{figure*}[htb]
\centering
\includegraphics[width=1.05\linewidth]{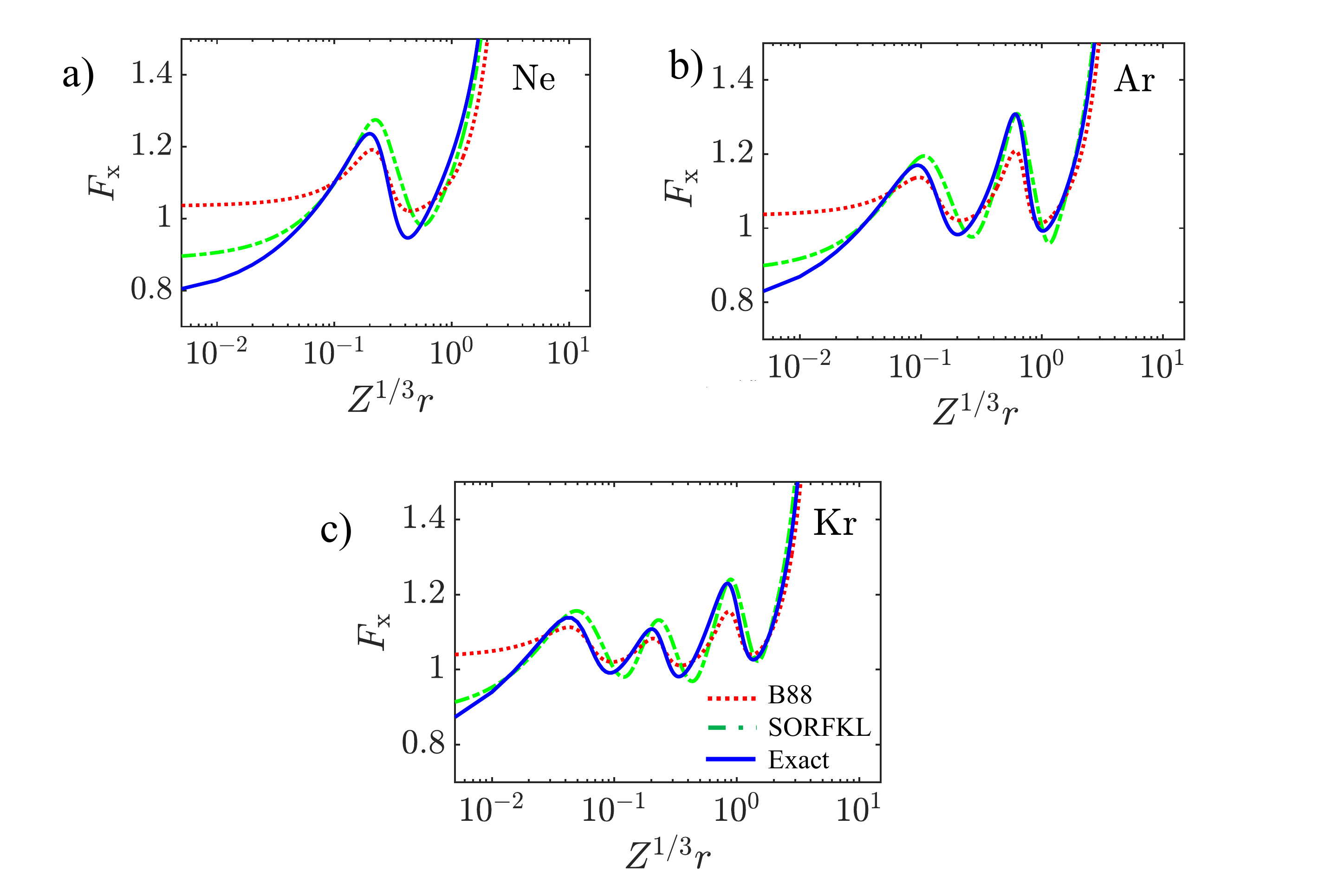}
\caption{Exchange enhancement factors of B88, SORFKL, and exact exchange in HG against $Z^{1/3}r$ (in Bohr) for Ne, Ar and Kr. Z is the nuclear charge.}
\label{fig:3}
\end{figure*}

Figure \ref{fig:3} shows the exchange enhancement factors in rare-gas atoms and complements Fig.~2 of the main text.

SORFKL's ability to recover the exact exchange HG energy density more closely than B88 stems from its mGGA nature. In particular, this allows SORFKL to recover the exact exchange HG energy density for both UEG and H, while maximally satisfying exact constraints. By incorporating the kinetic energy density via the dependence on $\beta$, SORFKL can distinguish different chemical environments~\cite{sun2013density}, allowing for the design of distinct exchange enhancement factors across these environments. For instance, \( \beta = 0 \) corresponds to single-orbital systems and regions such as H and core regions, while \( \beta \approx 1/2 \) characterizes slowly varying densities, enabling tailored GGA-like behavior for each regime~\cite{sun2015strongly}, as shown in Fig.~ \ref{fig:2}.

\section{Exchange enhancement factors and binding energy curves of $H_2^+$}
\label{H2Fx}

\begin{figure*}[htb]
    \centering
    \includegraphics[width=1\linewidth]{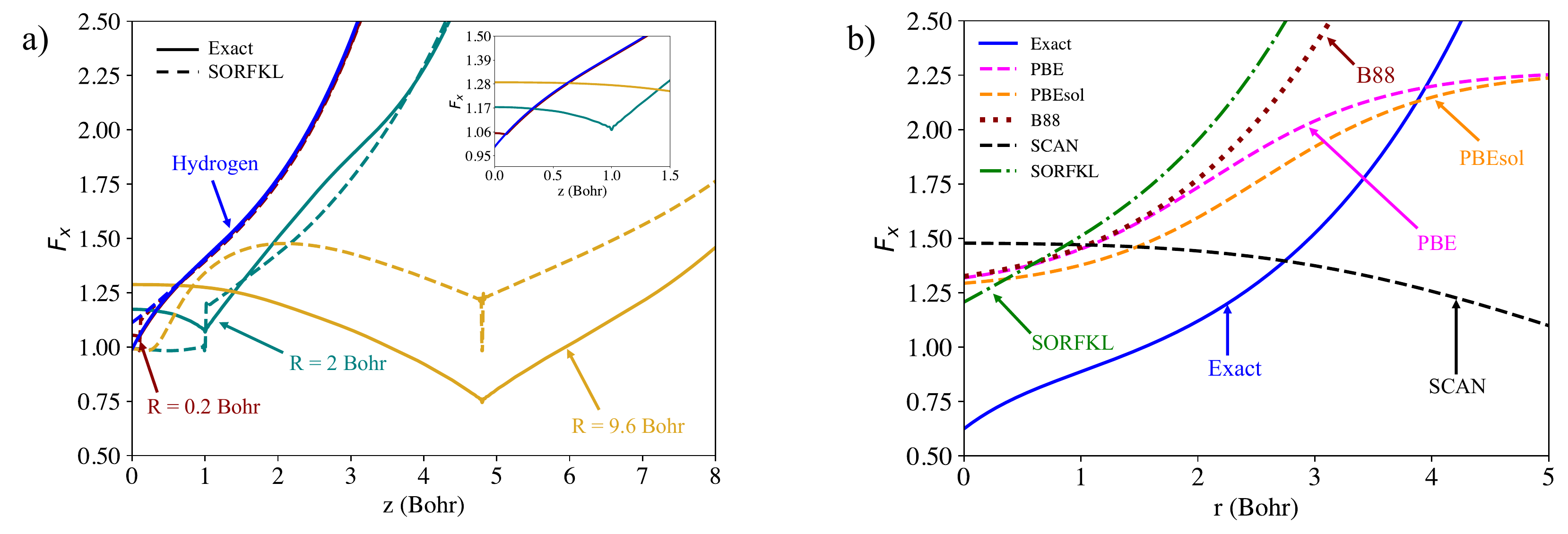}
    \caption{a) Exchange enhancement factors of exact exchange and SORFKL along the bond axis \(z\) of $H_2^+$ for various bond lengths \(R\), using the UGBS basis set. Solid lines represent the exact exchange \(F\x\), while dashed lines depict the \(F\x\) of SORFKL. The systems are centered at \(z=0\), with the nuclear positions of \(H_2^+\) for a given bond length \(R\) indicated by the cusps in \(F\x\) at \(z=R/2\). b) Enhancement factors of various DFAs compared with the exact exchange \(F\x\) for \(H_2^+(R\rightarrow \infty)\). At \(R\rightarrow \infty\), half of an electron remains bound to each proton, with the electron density \(n^H(r)/2\) around each proton. Since the bond center has negligible influence on the electron density near the protons, the exchange enhancement factors are computed using \(n^H(r)/2\) and plotted as a function of \(r\), where \(r=0\) represents the proton position. }
    \label{fig:exact_fx_h2+}
\end{figure*}

The \(\mathrm{H}_2^+\) molecular ion serves as a prototypical system for illustrating SIE, which arises from the incomplete cancellation of the spurious Coulomb interaction between an electron and itself. In \(\mathrm{H}_2^+\), the exchange hole is highly localized as the bond length \(R \to 0\) and becomes increasingly delocalized as \(R\) increases. In the limit \(R \to 0\), the exact exchange HG enhancement factor \(F\x^{\text{exact}}\) becomes that of hydrogen, whereas for \(R \to \infty\), it converges to \(F\x^{\text{exact}}\) of \(n^H(r)/2\) centered at each of the two protons, where \(n^H(r)\) is the electron density of an isolated hydrogen atom.

Figure~\ref{fig:exact_fx_h2+} a) illustrates the evolution of \(F\x\) with \(R\), showing that \(F\x^{\text{exact}}(z=0)\), the exact exchange HG enhancement factor at the bond center, decreases as $R$ decreases and converges to \(F\x^{\text{exact}}(r=0)\) at the nucleus of an isolated hydrogen as \(R \to 0^+\). In the limit \(R \to 0^+\), \(s=0\) remains at the bond center due to symmetry, while the electron density reduces to the hydrogenic \(1s\) orbital density of \(\mathrm{He}^+\), yielding an exact exchange HG enhancement factor identical to that of the isolated hydrogen atom. This confirms that \(F\x^{\text{exact}}(z=0)\) of \(H_2^+(R \to 0^+)\) equals that of the isolated hydrogen atom at its nucleus, justifying \(F\x^\textrm{SORFKL}(s=0, \beta=0) = F\x\Hyd(s_0)\).

As \(R\) increases, \(F\x^{\text{exact}}(z=0)\) grows and eventually diverges (\(F\x^{\text{exact}} \to \infty\) as \(R \to \infty\)), similarly to the two identical closed shell atoms discussed in the main text, and consistent with the expected asymptotic behavior of a finite system. Thus, for all $R$ we have $\beta=0$ and $s=0$ at the bond center, but $F\x^{\text{exact}}$ increases from $F\x\Hyd(s_0)$ for $R=0^+$ to infinity as $R\to\infty$. This underlines again the limitation of the semilocal form in recovering the exact exchange HG energy density everywhere. However, since the bond center in the dissociation limit is a region of delocalized exchange hole, a semilocal DFA such as SORFKL is not expected to be accurate there. Consequently, SORFKL closely matches the exact exchange HG enhancement factor for small \(R\), but the deviation increases with increasing \(R\).

Figure~\ref{fig:exact_fx_h2+} b) compares the enhancement factors of various DFAs with \(F\x^{\text{exact}}\) in HG for \(H_2^+(R\rightarrow \infty)\). In this dissociation limit, the bond center has negligible influence on the electron density around the protons, and the electron density around each atomic center becomes \(n^H(r)/2\). Consequently, the exact exchange HG enhancement factor around one nuclear center is given by \( 2^{-2/3} F\x^{\text{Hyd}}(r) \). For Figure~\ref{fig:exact_fx_h2+} b), we thus calculate the exchange enhancement factors using \(n^H(r)/2\) and plot them as a function of \(r\), with \(r=0\) representing the proton position. Apparently, the SORFKL exchange enhancement factor is approximately \( 2^{2/3} \) times larger than that of the exact exchange in HG, indicating that the underlying exchange hole of SORFKL is significantly deeper than the exact exchange hole for a reference electron at any spatial point, as expected. Similar behavior is observed for B88 with smaller $F\x$ than that of SORFKL, while PBE and PBEsol cross the exact exchange curve for \(r > 4\) Bohr. SCAN exhibits the strongest deviation from SORFKL and $F\x^\mathrm{exact}$ and crosses the exact exchange curve for \(r > 3\) Bohr. Correspondingly, the total exchange energies of \(H_2^+(R\rightarrow \infty)\), summed over both centers, are $-0.2405$ Ha (PBEsol), $-0.2540$ Ha (PBE), $-0.2588$ Ha (B88), $-0.2470$ Ha (SCAN), $-0.2700$ Ha (SORFKL), and $-0.15625$ Ha (Exact), the latter being $-E_\mathrm{H}=-5/32$ Ha, as required for a one-electron density.

\begin{figure}[htb]
    \centering
    \includegraphics[width=0.8\linewidth]{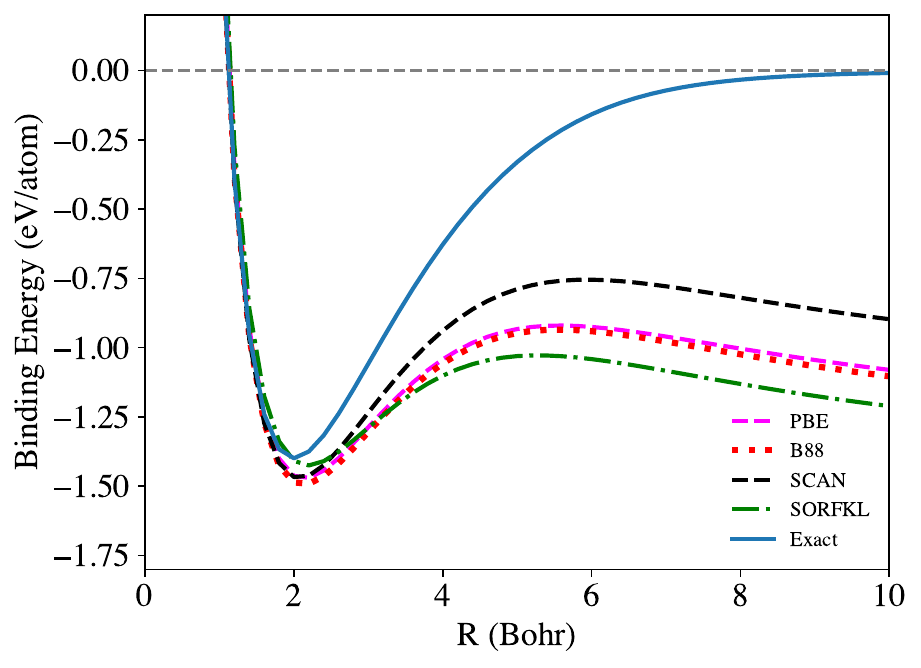}
    \caption{The binding energy curves of $H_2^+$ evaluated with Hartree-Fock(HF) and other exchange-only density functionals using the cc-pV5Z basis set.}
    \label{fig:be_h2+}
\end{figure}

As shown in Figure~\ref{fig:be_h2+}, at and before the equilibrium bond length (\(R = 1.058\) Å), the exchange energies of the semilocal DFAs align closely with the reference HF result. Interestingly, SORFKL provides the most accurate binding energy at equilibrium, outperforming all other DFAs considered, including SCAN. However, as the bond length increases and the exchange hole becomes more delocalized, all semilocal approximations yield increasingly negative binding energies, with SORFKL exhibiting the most pronounced deviation. 
This behavior is consistent with the semilocal DFA's performance in \(F\x\) and the associated exchange energies for \(H_2^+(R\rightarrow \infty)\), as illustrated in Figure~\ref{fig:exact_fx_h2+} b). These findings highlight the inherent limitations of semilocal density functionals in accurately describing \(H_2^+\) with stretched bonds, underscoring the necessity of incorporating nonlocal corrections into the functional form~\cite{vydrov2009nonlocal, ramasamy2025tackling}.

We emphasize once more that mGGAs are semilocal DFAs and thus not appropriate for modeling systems with delocalized exchange holes, where semilocal DFAs typically exhibit too negative exchange energies. Therefore, in the same gauge, semilocal density functionals are in general expected to have larger enhancement factors than the exact exchange for systems with delocalized exchange holes. This is what we observe for SORFKL in Figure~\ref{fig:exact_fx_h2+} a) for $R=9.6$ Bohr.

\clearpage

\begingroup\color{red}
\section{SORFKL combined with PBE correlation}
\label{sec:pbec}

To assess SORFKL in combination with an existing correlation functional, we pair SORFKL exchange with PBE correlation \cite{perdew1996generalized} (SORFKL+PBEc), evaluated non-self-consistently (see Sec.~\ref{sec:comp}) and compare it with exact (Hartree-Fock) exchange paired with the same correlation functional (HF+PBEc). Tables~\ref{tab:ae6}--\ref{tab:lc20} report atomization energies of the AE6 set \cite{lynch2003small}, binding energies of the S22 set of weakly interacting complexes \cite{jurecka2006benchmark}, forward ($V_f$) and backward ($V_b$) barrier heights~\cite{zheng2007representative} of three hydrogen-transfer reactions, and lattice constants of the LC20 set of solids \cite{sun2011self}.
SORFKL+PBEc improves over HF+PBEc for the AE6 atomization energies and the LC20 lattice constants, but is less accurate for the S22 binding energies and slightly less accurate for the barrier heights. These results show that SORFKL can be combined with an existing correlation functional; they are not intended to identify an optimal correlation partner for SORFKL, which is beyond the scope of this work. A correlation functional tailored to exchange in the Hartree gauge is under development and will be reported elsewhere.

\begin{table}[htb]
\color{red}
\caption{AE6 atomization energies (kcal/mol) from HF+PBEc and SORFKL+PBEc, compared with reference values.  The mean error and mean absolute error are defined as $\mathrm{ME}=\frac{1}{N}\sum_{i=1}^{N}(X_i^\mathrm{calc}-X_i^\mathrm{ref})$ and $\mathrm{MAE}=\frac{1}{N}\sum_{i=1}^{N}|X_i^\mathrm{calc}-X_i^\mathrm{ref}|$, respectively.}
\begin{ruledtabular}
\begin{tabular}{lccc}
Molecule & HF+PBEc & SORFKL+PBEc & Reference \\ \hline \hline
SiH$_4$          & 322.071  & 288.532  & 322.40  \\
S$_2$            & 83.012   & 112.152  & 101.67  \\
SiO              & 144.132  & 197.755  & 192.08  \\
C$_3$H$_4$       & 670.101  & 669.046  & 704.79  \\
C$_2$H$_2$O$_2$  & 558.258  & 624.588  & 633.35  \\
C$_4$H$_8$       & 1126.677 & 1066.581 & 1149.01 \\ \hline
ME               & $-33.175$   & $-24.108$   & --      \\
MAE              & 33.175   & 29.493   & --      \\
\end{tabular}
\end{ruledtabular}
\label{tab:ae6}
\end{table}

\begin{table}[htb]
\color{red}
\caption{S22 binding energies (kcal/mol) from HF+PBEc and SORFKL+PBEc, compared with reference values. Systems are numbered following the original S22 ordering \cite{jurecka2006benchmark}. Reference energies are the CBS[$\Delta$a(DT)Z] values of Ref.~\cite{takatani2010basis}. The mean error and mean absolute error are defined as $\mathrm{ME}=\frac{1}{N}\sum_{i=1}^{N}(X_i^\mathrm{calc}-X_i^\mathrm{ref})$ and $\mathrm{MAE}=\frac{1}{N}\sum_{i=1}^{N}|X_i^\mathrm{calc}-X_i^\mathrm{ref}|$, respectively.}
\begin{ruledtabular}
\begin{tabular}{rlccc}
No. & Complex & HF+PBEc & SORFKL+PBEc & Reference \\ \hline \hline
1  & (NH$_3$)$_2$                          & 2.5050  & 0.7305  & 3.17  \\
2  & (H$_2$O)$_2$                          & 5.0867  & 2.6640  & 5.02  \\
3  & Formic acid dimer                     & 21.7265 & 12.8861 & 18.80 \\
4  & Formamide dimer                       & 17.2866 & 9.7954  & 16.12 \\
5  & Uracil dimer (H-bonded)               & 21.7613 & 13.3221 & 20.69 \\
6  & 2-Pyridoxine$\cdot$2-aminopyridine    & 16.0677 & 10.2140 & 17.00 \\
7  & Adenine$\cdot$thymine (Watson-Crick)  & 15.5757 & 8.8321  & 16.74 \\
8  & (CH$_4$)$_2$                          & -0.1459 & -1.1902 & 0.53  \\
9  & (C$_2$H$_4$)$_2$                      & 0.5484  & -2.1523 & 1.50  \\
10 & Benzene$\cdot$CH$_4$                  & 0.3297  & -1.8710 & 1.45  \\
11 & Benzene dimer (parallel displaced)    & -1.4855 & -4.8088 & 2.62  \\
12 & Pyrazine dimer                        & 0.2917  & -4.2351 & 4.20  \\
13 & Uracil dimer (stacked)                & 5.4592  & -2.3699 & 9.74  \\
14 & Indole$\cdot$benzene (stacked)        & -1.6199 & -5.9922 & 4.59  \\
15 & Adenine$\cdot$thymine (stacked)       & 4.9186  & -4.5471 & 11.66 \\
16 & Ethene$\cdot$ethyne                   & 1.4819  & -0.3659 & 1.51  \\
17 & Benzene$\cdot$H$_2$O                  & 2.7709  & -0.1277 & 3.29  \\
18 & Benzene$\cdot$NH$_3$                  & 1.4739  & -1.1314 & 2.32  \\
19 & Benzene$\cdot$HCN                     & 4.8416  & 0.2338  & 4.55  \\
20 & Benzene dimer (T-shaped)              & 1.1857  & -2.3631 & 2.71  \\
21 & Indole$\cdot$benzene (T-shaped)       & 3.4394  & -1.1735 & 5.62  \\
22 & Phenol dimer                          & 5.4986  & 0.1808  & 7.09  \\ \hline
   & ME                                    & -1.4510 & -6.1086 & --    \\
   & MAE                                   & 1.9531  & 6.1086  & --    \\
\end{tabular}
\end{ruledtabular}
\label{tab:s22}
\end{table}

\begin{table}[htb]
\color{red}
\caption{Total energies (Ha) of reactants, products, and transition states (TS), and the resulting forward ($V_f$) and backward ($V_b$) barrier heights (kcal/mol) from HF+PBEc and SORFKL+PBEc, compared with reference barrier heights. The mean error and mean absolute error over the six barrier heights are defined as $\mathrm{ME}=\frac{1}{N}\sum_{i=1}^{N}(X_i^\mathrm{calc}-X_i^\mathrm{ref})$ and $\mathrm{MAE}=\frac{1}{N}\sum_{i=1}^{N}|X_i^\mathrm{calc}-X_i^\mathrm{ref}|$, respectively.}
\begin{ruledtabular}
\begin{tabular}{lccc}
Species & HF+PBEc & SORFKL+PBEc & Reference \\ \hline \hline
\multicolumn{4}{l}{Reaction 1: OH + CH$_4$ $\rightarrow$ CH$_3$ + H$_2$O} \\
OH      & -75.6950  & -75.7845  & -- \\
CH$_4$  & -40.5089  & -40.5185  & -- \\
CH$_3$  & -39.8324  & -39.8466  & -- \\
H$_2$O  & -76.3804  & -76.4819  & -- \\
TS      & -116.1689 & -116.3117 & -- \\
$V_f$   & 21.9211   & -5.3933   & 6.7 \\
$V_b$   & 27.5190   & 10.5491   & 20.2 \\ \hline
\multicolumn{4}{l}{Reaction 2: H + OH $\rightarrow$ O + H$_2$} \\
H       & -0.5053   & -0.5067   & -- \\
OH      & -75.6950  & -75.7845  & -- \\
O       & -75.0398  & -75.1149  & -- \\
H$_2$   & -1.1760   & -1.1730   & -- \\
TS      & -76.1743  & -76.2864  & -- \\
$V_f$   & 16.3593   & 3.0013    & 10.1 \\
$V_b$   & 26.0759   & 0.8858    & 13.1 \\ \hline
\multicolumn{4}{l}{Reaction 3: H + H$_2$S $\rightarrow$ H$_2$ + HS} \\
H       & -0.5053   & -0.5067   & -- \\
H$_2$S  & -399.3697 & -399.4168 & -- \\
H$_2$   & -1.1760   & -1.1730   & -- \\
HS      & -398.7197 & -398.7703 & -- \\
TS      & -399.8634 & -399.9243 & -- \\
$V_f$   & 7.3515    & -0.4953   & 3.6 \\
$V_b$   & 20.3160   & 11.9261   & 17.4 \\ \hline
ME      & 8.0738   & $-8.4377$   & -- \\
MAE     & 8.0738    & 8.4377    & -- \\
\end{tabular}
\end{ruledtabular}
\label{tab:bh}
\end{table}

\begin{table}[htb]
\color{red}
\caption{LC20 lattice constants (\AA) from HF+PBEc and SORFKL+PBEc, compared with reference values. The mean error and mean absolute error are defined as $\mathrm{ME}=\frac{1}{N}\sum_{i=1}^{N}(X_i^\mathrm{calc}-X_i^\mathrm{ref})$ and $\mathrm{MAE}=\frac{1}{N}\sum_{i=1}^{N}|X_i^\mathrm{calc}-X_i^\mathrm{ref}|$, respectively.}
\begin{ruledtabular}
\begin{tabular}{lccc}
Solid & HF+PBEc & SORFKL+PBEc & Reference \\ \hline \hline
Ag   & 4.1106 & 4.1029 & 4.063 \\
Al   & 3.9542 & 4.0759 & 4.019 \\
Ba   & 5.3723 & 4.9063 & 5.004 \\
C    & 3.4708 & 3.5932 & 3.555 \\
Ca   & 5.7593 & 5.4666 & 5.555 \\
Cu   & 3.6460 & 3.6213 & 3.595 \\
GaAs & 5.5132 & 5.7810 & 5.641 \\
Ge   & 5.5153 & 5.7983 & 5.644 \\
Li   & 3.5580 & 3.4391 & 3.451 \\
LiCl & 4.9960 & 5.2474 & 5.072 \\
LiF  & 3.8410 & 4.1514 & 3.974 \\
MgO  & 4.0799 & 4.3370 & 4.188 \\
NaCl & 5.5137 & 5.8296 & 5.565 \\
NaF  & 4.4673 & 4.8170 & 4.570 \\
Na   & 4.3113 & 4.1913 & 4.207 \\
Pd   & 3.7981 & 3.9309 & 3.876 \\
Rh   & 3.6327 & 3.8343 & 3.793 \\
Si   & 5.3456 & 5.5063 & 5.422 \\
SiC  & 4.2605 & 4.4223 & 4.348 \\
Sr   & 6.3825 & 5.9275 & 6.042 \\ \hline
ME   & -0.0028 & 0.0698 & -- \\
MAE  & 0.1251  & 0.1026 & -- \\
\end{tabular}
\end{ruledtabular}
\label{tab:lc20}
\end{table}

\endgroup

\clearpage
\bibliography{bib}